\documentclass[twocolumn]{elsart}
\usepackage{graphicx}
\usepackage{times,mathptmx}

\def\Journal#1#2#3#4{{#1} {\bf #2}, #3 (#4)}

\def\NIMA{{Nucl. Instr. Meth.} A}

\def\PRL{Phys. Rev. Lett.}

\def\APJ{Astrophys. J.}

\begin{document}
\begin{frontmatter}
\title{
Design of UHECR telescope 
with 1 arcmin resolution \\
 and 50 degree field of view
}


  \author[ICRR]{M.~Sasaki\thanksref{author}},
  \author[Tokyo]{A.~Kusaka} and
  \author[Tokyo]{Y.~Asaoka}

  \thanks[author]{E-mail:
    sasakim@icrr.u-tokyo.ac.jp}

  \address[ICRR]{Institute for Cosmic Ray Research, University of Tokyo,
    5-1-5 Kashiwanoha, Kashiwa, 277--8582 Japan}
  \address[Tokyo]{Department of Physics, University of Tokyo,
    7-3-1 Hongo, Bunkyo, Tokyo, 113--0033 Japan}


  \begin{abstract}
A new telescope design based on Baker-Nunn optics is proposed for
observation of ultra high-energy cosmic rays (UHECRs).
The optical system has an image resolution of smaller than
0.02$^\circ$ within a wide field of view of 50$^\circ$ angular diameter.
When combined with a high-quality imaging device,
the proposed design enables the directions of UHECRs
and high-energy neutrinos 
to be determined with accuracy of better than 1 arcmin.
%
%
%
The outstanding resolution of this telescope allows charge-separated
cosmic-rays to be resolved and the source to be determined accurately.
This marked improvement in angular resolution will 
allow the multi-wavelength and ``multi-particle'' observations
of astronomical objects
through collaboration with established astronomical observations.
%
  \end{abstract}
\end{frontmatter}

\setlength{\oddsidemargin}{-1cm}
\setlength{\evensidemargin}{-1cm}

\section{Introduction}
The observation of ultra high-energy cosmic rays (UHECRs) and
high-energy neutrinos (HE-$\nu$'s) with directional accuracy of
$\leq 1$ arcmin (') is expected to lead to revolutionary progress in
particle astrophysics through ``multi-particle'' observations 
in cooperation with multi-wavelength astronomical observations.
%
An excellent example is the recent prompt observation of X-rays from a
gamma-ray burst (GRB) with directional precision of 1' by Beppo-SAX,
an event that marked the start of multi-wavelength observations using
many advanced astronomical telescopes \cite{Beppo1,Beppo2,Beppo3}.
The directional accuracy of 1' is several tens times better than 
that of $\sim1^\circ$ currently operating, constructing, or planning
UHECR detectors \cite{AGASA,HiRes,Auger,TA} aim for.

\vspace*{0.4cm}
Upon entering the atmosphere, UHECRs produce an air shower (AS)
through interaction with air nuclei.
The AS axis coincides with the UHECR incident angle,
and the intrinsic angular resolution of the shower is 0.3' or smaller
for a UHECR of 10$^{17}$~eV. 
This narrow dispersion results from the high Lorenz factor of secondary
particles ($\sim 10^4$) in the direction of the parent particle, and 
the very low transverse momenta of secondary particles (rarely
exceeding 1~GeV/c) as reported by the ATLAS group \cite{TDR_ATLAS}. 
For UHECRs, more precise reconstruction of the primary direction can
be expected due to the higher Lorenz factor. 

ASs can be detected via the isotropic fluorescence
emission of the air nitrogen that it excites.
An optical system with imaging devices can be used to track the
longitudinal development of an AS. 
This technique was first employed in the Fly's Eye detector~\cite{FlysEye} 
and will be part of several future projects on UHECR detection.
The key advantage of fluorescence detection is the capacity for
achieving fine angular resolution. 
The signal to noise ratio ($S/N$) for each pixel of the imaging
devices is inversely proportional to the square root of the angular
diameter of the field of view (FOV).
A better $S/N$ will allow the device to trigger charge levels, making
it possible to analyze less energetic or more distant events. 

Galactic and extragalactic magnetic fields are considered to have
intensities on the order of $\mu$G, nG, and below \cite{Kronberg}.
Thus, UHECRs are expected to experience only slight deflection due to
galactic 
and extragalactic magnetic fields, and  cannot be confined by the galactic
magnetic field. 
The existence of an ``ankle'' at $E \sim 10^{19}$~eV is often interpreted
as a crossover from a steeper galactic component to a harder
component of extragalactic origin.

As the possible sources of astrophysical accelerators capable of
producing UHECRs are very limited,
UHECRs that have been deflected only slightly can be traced back
to their sources.
Thus, high-resolution measurements of the arrival direction will
lead to significant progress in cosmic-ray astronomy. 
The magnetic deflection of UHECRs during propagation induces
charge separation.
One of the important implications of 1' angular resolution is that 
this separation of charge can be resolved, allowing
unambiguous particle identification to be performed based on the
deflection from sources.
Such observations may also allow us to study the
structure of extragalactic magnetic fields.

Based on the above considerations, we
propose a telescope design that features a FOV of 50 $^\circ$ $\times$
50 $^\circ$ and 1' resolution for the arrival direction.
This is realized through a combination of a large-area
image intensifier (II) connected to a CCD camera~\cite{II-CCD} and 
a customized Baker-Nunn optical system.
In this paper, we describe the conceptual design of the
modified Baker-Nunn optical system.
The expected performance of the telescopes for UHECR
detection is also presented assuming that the image resolution of the
CCD camera is better than that of the optical system. 

\section{General Concept of Optical System}

Our design requirements can be summarized as follows:
\begin{enumerate}
\item a large FOV of 50$^\circ$ $\times$ 50$^\circ$ to cover a large solid 
angle for UHECR detection;
\item an image resolution of $\sim$0.02$^\circ$ to achieve 1'
resolution for the arrival direction;
\item a large caliber with a diameter of $\sim$2000~mm to
distinguish faint fluorescence.
\end{enumerate}
In addition, cost performance will be a critical issue with respect to
construction of a large-aperture array of telescopes.

Among several approaches to meet our requirements,
spherical-mirror optics with an aperture diaphragm on the
mirror's curvature center appears to offer some advantages
for the proposed telescope. 
This simple optics produces
homogeneous images within the entire FOV.
Typical off-axis aberrations,
principally coma aberration, are eliminated,
although spherical aberration is a factor.
Thus, from a review of the variants of this optics,
the Schmidt, Makstov, Baker Super-Schmidt, and Baker-Nunn optics were
selected as candidates.
The Schmidt optics has a one-element corrector lens at the diaphragm,
and provides higher resolution than optics without the corrector lens.
However, we found that it was difficult to achieve the required
image resolution over the wide FOV.
The Makstov and Baker Super-Schmidt optical systems require thick
corrector lenses, and as such are not suited to large-scale
telescopes.

\begin{figure}[hbtp]
 \begin{center}
  \includegraphics[width=80mm]{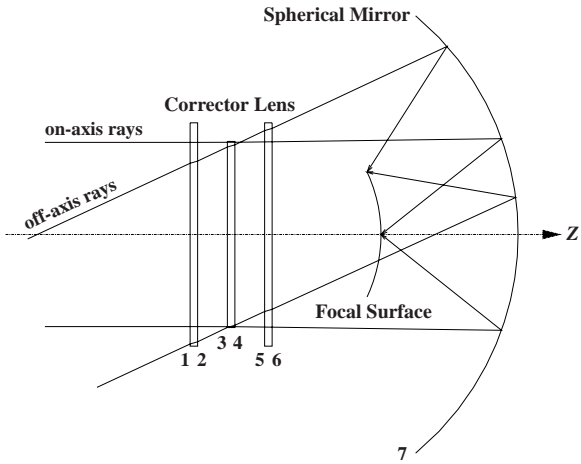}
  \caption{Schematic of proposed telescope based on
  Baker-Nunn optics.
  The second lens element (3,4) is an aperture diaphragm.
  }
  \label{fig:optics}
 \end{center}
\end{figure}

The Baker-Nunn optics has a three-element corrector lens 
consisting of convex and concave lenses with spherical or
aspherical surfaces~\cite{Baker-Nunn}.
The basic premise of this optics is that
each element has easy-to-make spherical surfaces,
residual spherical aberration after assembly can be 
canceled by adjusting the spacing between lens elements, and
chromatic aberration can be eliminated by selecting appropriate lens 
materials. 
Figure~\ref{fig:optics} shows a schematic of the Baker-Nunn
optics. 
The second lens element is an aperture diaphragm.

Although the advantages of the Baker-Nunn optics are not related
directly to the specific area of improvement pursued in the proposed
design, the three-element corrector lens in this optics provides a
large degree of freedom for optimization.
As a result, it is considered the most suitable for our purposes.
In our design, the first and third lens elements are
2400~mm in diameter, and the second element is 2000~mm in diameter,
and all elements are 10~mm thick.
Considering the cost, and complexity of fabrication processes,
we designed the corrector lens as an acrylic Fresnel lens as
appropriate for a large-aperture telescope.
As described in the following section, the surface shape of each lens
element was determined through an optimization process.

\section{Optimization by Ray Tracing}
\subsection{Ray Tracing}
We developed ray-tracing program written in C++ in order to optimize
the optical system and examine its performance.
The program models the refraction and reflection of incident rays on each
surface, and allows a bundle of on- and off-axis rays to be 
traced numerically.
As is common in designing an optical system, the shape of the front
surface of each optical component is expressed as
\begin{equation}
 Z = \frac{h^2 / R}{1 + \sqrt{1 - h^2 / R^2}}
  + A h^4 + B h^6 + C h^8,
\end{equation}
where $Z$, $h$, and $R$ represent the surface sag, radial
height, and curvature of the surface.
$Z$ is positive in the direction of on-axis incident
rays.
$A$, $B$, and $C$ represent the 4th, 6th, and 8th order
aspheric coefficients.
The surfaces of each lens element and the mirror are numbered 1
to 7 as shown in Fig.~\ref{fig:optics}.

The focal surface is defined as the surface minimizing
the image spot for a bundle of parallel rays.
The spot size is defined as the root mean square (RMS) of positions of
parallel rays at the focal surface.
The fineness of the bundle of rays analyzed is determined according to
the purpose of ray tracing: a rather rough interval between rays is
selected for optimization, and a fine interval is selected for
performance evaluation.
The spot size ($s$) is a function of the incident angle ($\theta$) and
the wavelength ($\lambda$) of parallel rays.

\subsection{Optimization}

We used the modified Powell's method as the algorithm for
optimization based on the following evaluation function:
\[
F = \sum_\theta \, ( \sum_\lambda \, s(\theta, \lambda)^2 \, )^2,
\]
where $\theta$ varies from 0$^\circ$ to 25$^\circ$,
and $\lambda$ varies from 330 to 410~nm,
taking into account the air-fluorescence emission spectrum.
We adopted the fourth power of spot size as the evaluation
function in order to obtain flat resolution over the entire FOV.

In principle, the curvature, aspheric coefficients, thickness, and position of
each lens element are free parameters in optimization.
In order to reduce the number of free parameters, we imposed the
symmetry described in Ref.~\cite{Baker-Nunn} in which the corrector
lens is symmetric with respect to the plane including the center
of the second element and which is perpendicular to the axis of the ray.
In addition, the thickness and position of each surface were fixed.
These constraints allowed much more efficient optimization.
The optimized parameters of the optical system are given in
Table~\ref{tab:optim-params}.
The axis and origin of the $Z$ coordinate are coincident with the axis of
the rays and the center of the corrector lens.
Note that we took the effect of the region obscured by
the focal surface into account in the optimization.
\begin{table}[hbtp]
\renewcommand{\arraystretch}{1.25}
\caption{
Optimized parameters of the optical system.
}
\label{tab:optim-params}
\begin{center}
 \begin{tabular}{lrrc}
  \hline \hline
  \multicolumn{1}{c}{Surface} & \multicolumn{1}{c}{Z}
   & \multicolumn{2}{c}{Parameters} \\
  \hline
  1: & $-405.0$ & \hspace{5mm} R: & $\infty$ \\
  \hline
  2: & $-395.0$ & R: & $-14767.2$ \\
  && A: & $2.30448 \times 10^{-11}$ \\
  && B: & $-7.03756 \times 10^{-18}$ \\
  \hline
  3: & $-5.0$ & R: & $-30479.6$ \\
  && A: & $1.45344 \times 10^{-11}$ \\
  && B: & $-6.37472 \times 10^{-18}$ \\
  && C: & $-8.78219 \times 10^{-25}$ \\
  \hline
  4: & $5.0$ & R: & $30479.6$ \\
  && A: & $-1.45344 \times 10^{-11}$ \\
  && B: & $6.37472 \times 10^{-18}$ \\
  && C: & $8.78219 \times 10^{-25}$ \\
  \hline
  5: & $395.0$ & R: & $14767.2$ \\
  && A: & $-2.30448 \times 10^{-11}$ \\
  && B: & $7.03756 \times 10^{-18}$ \\
  \hline
  6: & $405.0$ & R: & $\infty$ \\
  \hline
  7: & $3082.0$ & R: & $-3060.0$ \\
  \hline \hline
  \multicolumn{4}{r}{(unit: mm)}
 \end{tabular}
\end{center} 
\end{table}

\begin{figure}[hbtp]
 \begin{center}
  \includegraphics[width=68mm]{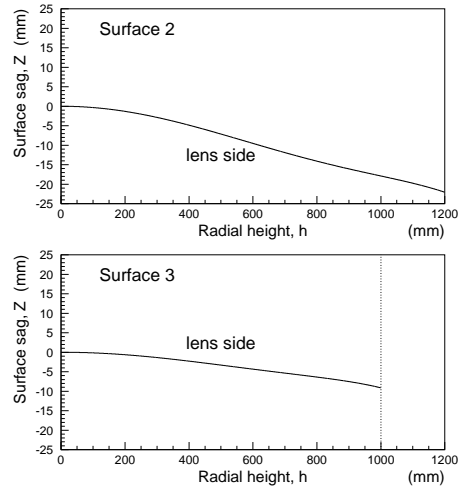}
  \caption{Sag of Surfaces~2 and 3 as a function of radial 
  height.
  The lens surface is constructed from a series of concentric grooves 
  (Fresnel lens), each 10-mm thick.
  }
  \label{fig:sag}
 \end{center}
\end{figure}

Figure~\ref{fig:sag} shows the surface sag of lenses 2, and 3 as a
function of radial height.
The surface sag remained well below 0.5~mm when the width of Fresnel
grooves was chosen to be 10 mm.
Thus, the 10-mm thickness of each lens element is sufficient, although the
thickness may be changed according to the support method employed for 
the corrector lens.
We confirmed that the same level of spot-size resolution can be obtained by
another optimization using a different lens thickness.

\begin{figure*}[hbtp]
 \begin{center}
  \includegraphics[width=135mm]{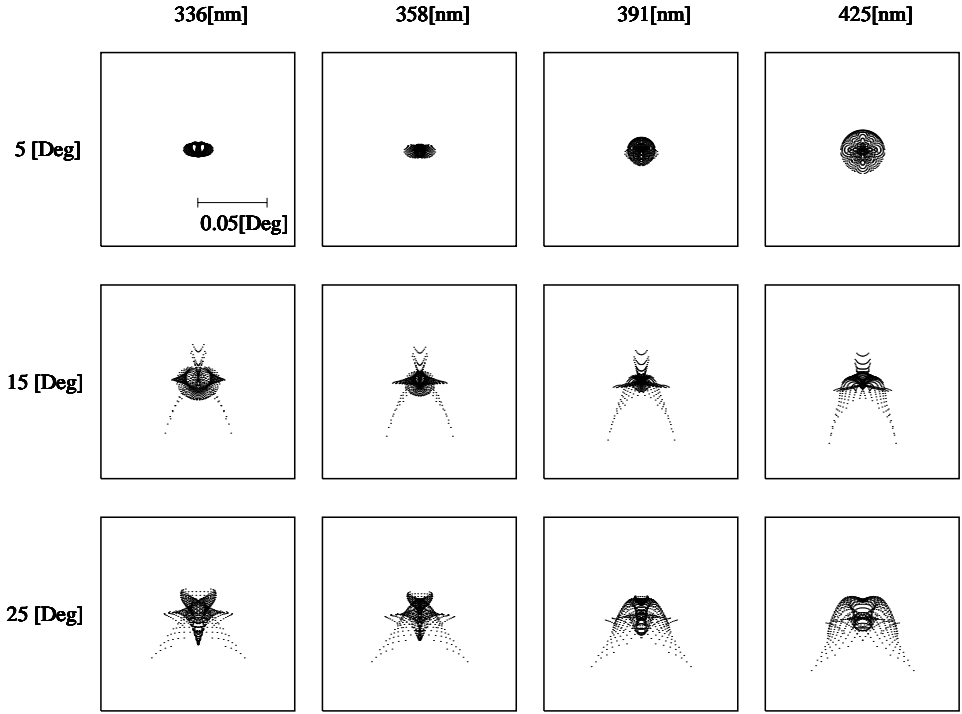}
  \caption{Spot diagrams for optimized optical system.}
  \label{fig:spot}
 \end{center}
\end{figure*}

Figure~\ref{fig:spot} shows the spot diagrams of the optimized optical 
system. 
The shape and size of spot diagrams changed with the wavelength and
the incident angle of the incoming rays.
The four strongest fluorescence peaks, $\lambda=$~336,~358,~391, and~425~nm,
which dominate the filtered fluorescence spectrum after propagation
through air, were selected as wavelengths for the diagrams. 
The spot diagrams for incident angles of 5$^\circ$, 15$^\circ$, and
25$^\circ$ are shown.

\begin{figure}[hbtp]
 \begin{center}
   \includegraphics[width=68mm]{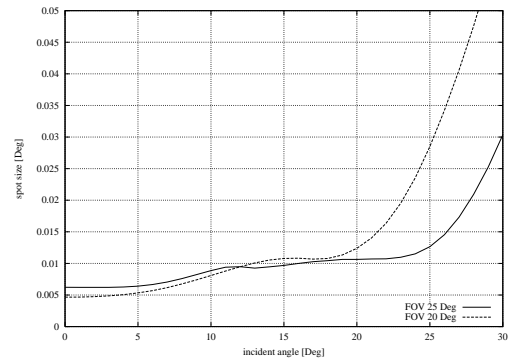}
  \caption{RMS distribution of spot size as a function of incident 
   angle.
  }
 \label{fig:reso}
 \end{center}
\end{figure}

Figure~\ref{fig:reso} shows the spot size distribution as a
function of incident angle, taking the characteristic wavelengths 
in the fluorescence spectrum into account.
Solid and dashed curves represent the spot size optimized to cover a
FOV of $\pm$25$^\circ$ and $\pm$20$^\circ$, respectively.
The lens parameters shown in Table~\ref{tab:optim-params} are the
result of the former optimization,
and give a spot size of smaller than
0.02$^\circ$ over a FOV of $\pm$25$^\circ$.
The results were confirmed by cross checks using Mathematica Optica
\cite{MathOpt} with correction for the refractive index of the
atmosphere.

In parallel with the development of this optical system, the present
authors have also developed a high-density CCD imaging detector with
integrated large-area image intensifier, which has been
demonstrated to have a resolution of $\sim$ 0.02$^\circ$ \cite{II-CCD}.
This imaging system is mounted at the focal surface of the optimized
optics to give the proposed telescope design.

\section{Performance}
A solid angle of nearly 2$\pi$ sr can be observed using an array of 8
telescopes (i.e., a station), with 7 telescopes arranged
azimuthally and 1 telescope directed upward.
It should be noted that the telescope directed upward is primarily
important in HE-$\nu$ detection rather than UHECR detection
\cite{nu_TA}. 

A Monte Carlo simulation was performed to examine the resolution
obtained for the arrival direction of UHECRs.
The event generator calculated the signal strength
produced by atmospheric fluorescence emission
from ASs with a given geometry, and built
simulation events by calculating the number of photoelectrons
detected by each pixel of a CCD camera at each mirror.
The simulation includes the longitudinal and lateral profiles of the
AS, 
the model of the atmosphere, the air fluorescence yield, and the
detector optics.
We also took into account fluctuation of the first interaction depth,
impact point, and directional angles of shower cores assuming
appropriate distributions.

\begin{figure}[hbtp]
 \begin{center}
  \includegraphics[width=68mm]{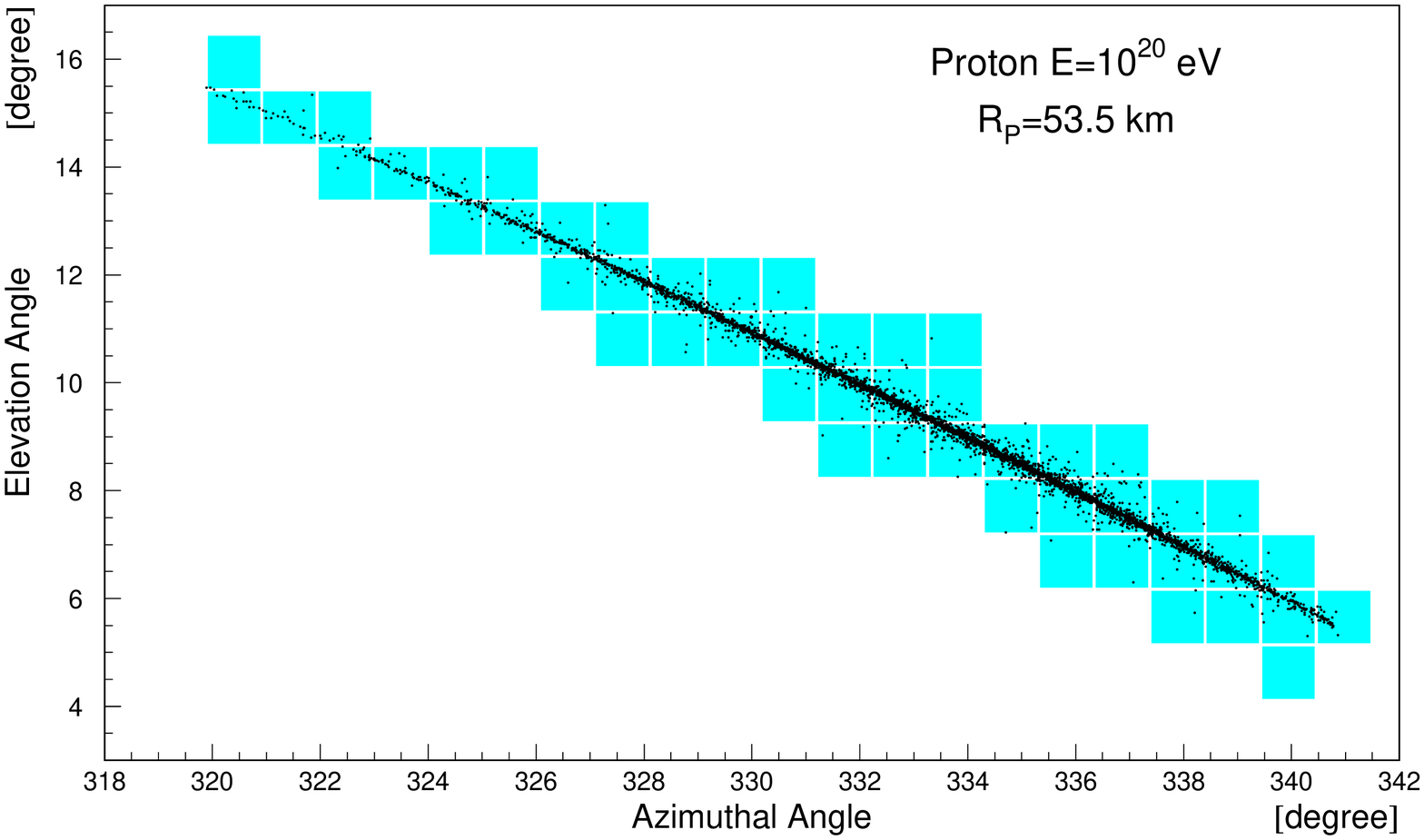}
  \includegraphics[width=68mm]{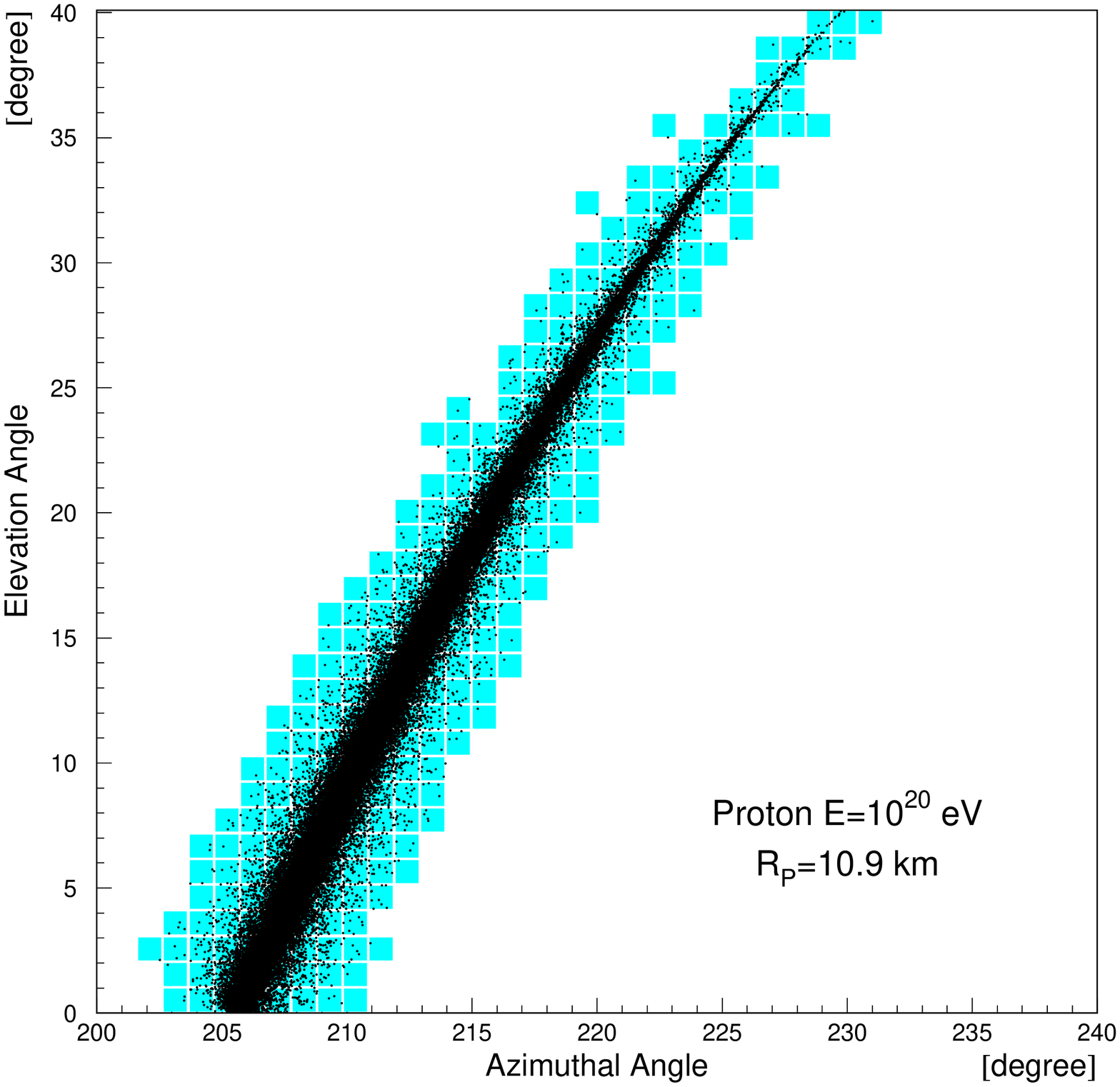}
 \caption{Two examples of AS events for 10$^{20}$ eV protons,
 showing events with impact parameter of $R_{\rm P}$ = 53.5 and 10.9 km.
 }
 \label{fig:event}
 \end{center}
\end{figure}

Current and planned fluorescence detectors \cite{HiRes,TA}
have a pixel resolution of $\sim$1$^\circ$.
The angular resolution of the reconstructed track
was evaluated by analyzing event samples that triggered
the current 1$^\circ$-pixel detectors.
Such signals are triggered by having at least 6 pixels ($1^\circ$
resolution) fire at 4$\sigma$ greater than the night sky background,
with a corresponding reconstructed track extending more than 5$^\circ$ 
in the FOV \cite{HiRes,TA}.
A night sky background level of 100
photoelectrons/$\mu$s/($1^\circ \times 1^\circ$) was assumed.
Figure~\ref{fig:event} shows the comparison of event displays between
the 0.02$^\circ$ and 1$^\circ$ systems.

\begin{figure}[hbtp]
 \begin{center}
  \includegraphics[width=68mm]{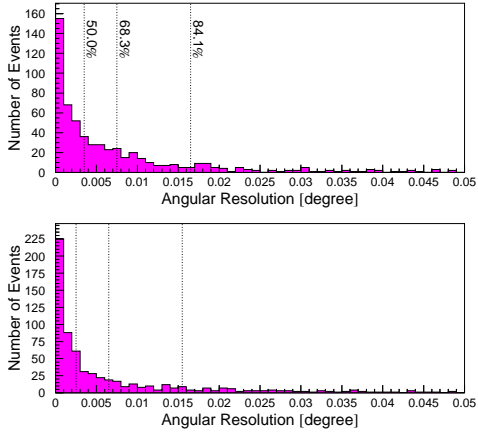}
  \caption{The angular resolution of SDP for (Top) 10$^{19}$ and
  (Bottom) 10$^{20}$~eV protons.}
\label{fig:angle}
 \end{center}
\end{figure}

\begin{figure}[hbtp]
 \begin{center}
  \includegraphics[width=68mm]{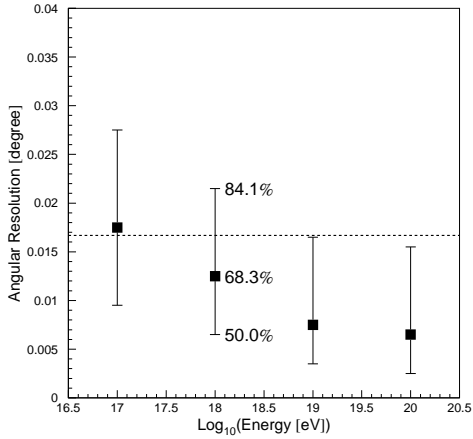}
  \caption{The energy dependence of the angular resolution for SDP
  determination.}
\label{fig:evsang}
 \end{center}
\end{figure}

Applying a stereo reconstruction of the AS,
the arrival direction is determined by the geometrical
intersection of shower-detector planes (SDPs).
Since the resolution of the normal vector of a SDP is the primary
factor affecting the resolution of arrival direction,
we discuss here for simplicity the accuracy of SDP measurement.
The angular resolution of SDP is
plotted in Fig.~\ref{fig:angle} for 10$^{19}$ and 10$^{20}$~eV
protons.
The angular resolution is determined from the integral of the
distribution encompassing 68~\% of events.
Figure~\ref{fig:evsang} shows the variation in the angular
resolution of the SDP with proton energy.
Resolution of $<$~1' is obtained at above 10$^{18}$~eV for the SDP.
In the case of stereo reconstruction with two or more stations, the
pointing resolution is also expected to be $\sim$~1' after 
reconstruction of the geometrical intersection of SDPs.

\begin{figure}[hbtp]
 \begin{center}
  \includegraphics[width=68mm]{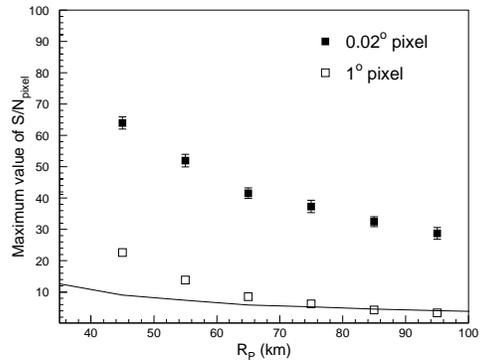}
 \caption{Averaged maximum values of $S/N$ for 0.02$^\circ$ and
  1$^\circ$ pixels as a function of $R_{\rm P}$ for 10$^{20}$~eV
  protons.} 
 \label{fig:snmax}
 \end{center}
\end{figure}

The $S/N$ of each pixel is proportional to
$( A/\Delta \theta )^{\frac{1}{2}}$ \cite{FlysEye},
where $A$ represents the light collection area, corresponding to the
reflection area 
for simple mirror optics and the area of the aperture diaphragm for
the Baker-Nunn optics,
and $\Delta \theta$ denotes the angular size of each pixel.
Note that this proportionality holds only when the effects of lateral
distribution are neglected.
Assuming the same size of $A$,
the $S/N$ value for the 0.02$^\circ$-pixel imaging device
is expected to be about 7 times higher than that from the 1$^\circ$-pixel
one. 
However, this is expected to be affected by the 
quantized number of photoelectrons at these small dimensions.
To examine this, the AS profile was simulated by the Monte Calro
method and the quantitative improvement in the $S/N$ was examined
for the 0.02$^\circ$ pixels.
The AS events of protons with energy of 10$^{20}$~eV were
generated isotropically for a given impact parameter ($R_{\rm P}$). 
The maximum value of $S/N$ (obtained event-by-event) was averaged for
a given $R_{\rm P}$,
and the maximum values are shown in Fig.~\ref{fig:snmax}.
Note that the event selection applied here differs from that applied to
compare the angular resolution
as it is not necessary to apply the 6 pixel condition to
compare the $S/N$ of the $1^\circ$- and 0.02$^\circ$-pixel detectors
for distant ASs. 
%
%
The solid curve shows the expected $S/N$ value for 1$^\circ$ pixels
calculated from that of 0.02$^\circ$ pixels by the theoretical
proportionality.
The expected and obtained $S/N$ values are in good agreement for the distant
events ($R_{\rm P} >$ 70 km). 
For the closer events, the lateral distribution of the shower-core
image becomes significant compared with that for the 1$^\circ$ pixel
image (see the bottom panel of Fig.~\ref{fig:event}). 
As a result, the expected $S/N$ ratio is appreciably lower than the
simulation result.

\begin{figure}[hbtp]
 \begin{center}
  \includegraphics[width=68mm]{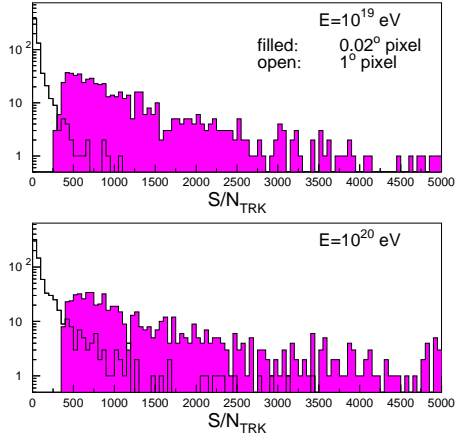}
 \caption{Comparison of $S/N_{\rm TRK}$ for 0.02$^\circ$ and
  1$^\circ$ devices for 10$^{19}$ and 10$^{20}$~eV
  protons.} 
 \label{fig:sn_trk}
 \end{center}
\end{figure}

We also define another signal to noise ratio ($S/N_{\rm TRK}$) appropriate
for the track captured by the imaging devices.
The signal ($S$) is the sum of charges detected by all pixels
associated with the track.
The background ($BG$) is the sum of background for all pixels
associated with the track.
Then, $S/N_{\rm TRK}$ is defined as $S/N_{\rm TRK}=S/\sqrt{BG}$.
Figure~\ref{fig:sn_trk} shows the distributions of $S/N_{\rm TRK}$ for
protons of 10$^{19}$ and 10$^{20}$~eV.
The 0.02$^\circ$ device exhibited excellent $S/N_{\rm TRK}$ values.
The significant improvement in $S/N_{\rm TRK}$ was
obtained because the lateral extent of the AS is smaller than the pixel
resolution of 1$^\circ$.
Thus, we have demonstrated the improved resolution of the arrival
direction offered by the proposed system.
The detailed performance of the detector, including aperture, energy
resolution, and shower-max resolution, will be discussed
elsewhere. 

\section{Conclusion}
We have presented the conceptual
design of a new type of telescope for UHECR detection based on
the Baker-Nunn optical system.
The telescope has a spot size resolution of smaller than 0.02$^\circ$
within the entire FOV of 50$^\circ$, and has been shown to be capable
of 1' angular resolution for the arrival direction of UHECRs.
This design has the potential to open new possibilities in
the field of UHECR and HE-$\nu$ physics through high-resolution
observation.

\begin{ack}
The authors express their gratitude to Y. Higashi for helpful advice
and fruitful discussion, and to T. Aoki for continuous support.
We also thank N. Manago and M. Jobashi for their useful help
in the development of the Monte Carlo simulation.
\end{ack}

\end{document}